\documentclass[reprint,
superscriptaddress,
amsmath,amssymb,aps,prl,floatfix]{revtex4-2}
\usepackage[dvipsnames]{xcolor}
\usepackage{graphicx}
\usepackage{dcolumn}
\usepackage{bm}
\usepackage[mathlines]{lineno}
\usepackage{hyperref}
\hypersetup{colorlinks=true,linkcolor=red,citecolor=blue,urlcolor=magenta}
\usepackage{graphicx}
\usepackage{multibib}
\usepackage{braket}

\begin{document}

\title{Realizing quantum speed limit in open system with a PT-symmetric trapped-ion qubit}

\author{Pengfei Lu}
\thanks{These three authors contribute equally}
\affiliation{School of Physics and Astronomy, Sun Yat-Sen University, Zhuhai, 519082, China}
\author{Teng Liu}
\thanks{These three authors contribute equally}
\affiliation{School of Physics and Astronomy, Sun Yat-Sen University, Zhuhai, 519082, China}
\author{Yang Liu}
\thanks{These three authors contribute equally}
\affiliation{School of Physics and Astronomy, Sun Yat-Sen University, Zhuhai, 519082, China}
\affiliation{Center of Quantum Information Technology, Shenzhen Research Institute of Sun Yat-sen University, Nanshan Shenzhen 518087, China}
\author{Xinxin Rao}
\affiliation{School of Physics and Astronomy, Sun Yat-Sen University, Zhuhai, 519082, China}
\author{Qifeng Lao}
\affiliation{School of Physics and Astronomy, Sun Yat-Sen University, Zhuhai, 519082, China}
\author{Hao Wu}
\affiliation{School of Physics and Astronomy, Sun Yat-Sen University, Zhuhai, 519082, China}
\author{Feng Zhu}
\affiliation{School of Physics and Astronomy, Sun Yat-Sen University, Zhuhai, 519082, China}
\affiliation{Center of Quantum Information Technology, Shenzhen Research Institute of Sun Yat-sen University, Nanshan Shenzhen 518087, China}
\author{Le Luo}
 \email{luole5@mail.sysu.edu.cn}
\affiliation{School of Physics and Astronomy, Sun Yat-Sen University, Zhuhai, 519082, China}
\affiliation{Center of Quantum Information Technology, Shenzhen Research Institute of Sun Yat-sen University, Nanshan Shenzhen 518087, China}
\affiliation{International Quantum Academy, and Shenzhen Branch, Hefei National Laboratory, Futian District, Shenzhen, 518017, China}

\begin{abstract}

Evolution time of a qubit under a Hamiltonian operation is one of the key issues in quantum control, quantum information processing and quantum computing. It has a lower bound in Hermitian system, which is limited by the coupling between two states of the qubit, while it is proposed that in a non-Hermitian system it can be made much smaller without violating the time-energy uncertainty principle. Here we have experimentally confirmed the proposal in a single dissipative qubit system and demonstrate that the evolution time of a qubit from an initial state to an arbitrary state can be controlled by tuning the dissipation intensity in a non-Hermitian Parity-Time-Symmetric ($\mathcal{P T}$-symmetric) quantum system. It decreases with increasing dissipation intensity and also gives a tighter bound for quantum speed limit (QSL). We also find that the evolution time of its reversal operation increases with the increasing dissipation intensity. These findings give us a well-controlled knob for speeding up the qubit operation, and pave the way towards fast and practical quantum computation, opening the door for solving sophisticated problems with only a few qubits.

\end{abstract}

\maketitle


\noindent\textit{\textbf{Introduction}}\quad Quantum computation \cite{benioff1980computer,
feynman1982simulating} is the emerging crossroad connecting mathematical computation and quantum mechanics.~Trapped ions, superconducting circuits, photons, nitrogen-vacancies, neutral atoms in an optical lattice, polar molecules, quantum dot, have been proposed as candidates for quantum computation. Despite many promising successes, there are a number of technical challenges
\cite{divincenzo2000physical}, including scaling up the number of qubits to a formidable amount for general computation task. However, as the number of qubits grows,
the gate speed usually slows down with $\frac{1}{\sqrt{N}}$ \cite{brown2016co}, leading to the decoherence resulted from the coupling between the qubits and the
surrounding environment. In order to perform a complex computational task within the coherence time, it would be preferential to speed up the qubit operational time as fast as possible.

In traditional quantum computing, the qubit is a two-level system described by a Hermition Hamiltonian,  in which the gate operation time is limited to a so-called quantum speed limit (QSL). This limit describes the  unitary evolution time of a qubit from an initial state $|\psi_i\rangle$ to a target state $|\psi_f\rangle$, which is bounded by $\tau_f=\max(\tau_{MT},\tau_{ML})$, where the Mandelstam-Tamm limit $\tau_{MT}=\pi \hbar/2\Delta E$ \cite{mandelstam1991uncertainty, ness2021observing} and the Margolus-Levitin limit $\tau_{ML}=\pi \hbar/2\langle E\rangle$ \cite{margolus1998maximum, ness2021observing} relate the maximum speed of evolution to the system’s energy uncertainty and mean energy, respectively.

In recent years, substantial efforts have been implemented to speed up the qubit operational time by reducing the QSLs \cite{pfeifer1993fast, margolus1998maximum, levitin2009fundamental}. Meanwhile, instead of Hermitian system, QSL in an open system described by non-Hermitian Hamiltonian has become the focus of intensive theoretical studies for its intriguing potential towards practical quantum computation \cite{mostafazadeh2007quantum,gunther2008p,gunther2008naimark,assis2008quantum, taddei2013quantum, del2013quantum,deffner2013quantum,marvian2015quantum,funo2019speed}. Specifically, for a non-Hermitian $\mathcal{PT}$-symmetric system, it has been proposed that the qubit operational time can approach to an infinitesimal time scale without violating the time-energy uncertainty principle \cite{bender2007faster, brody2019evolution}, yet to be realized in a qubit architecture compatible with universal quantum computation.

Trapped ions are one of the most promising platforms for universal quantum computing, demonstrating the single- and two-qubit gate fidelities in small-scale systems necessary for building a scalable system combined with the ions' long coherence times \cite{debnath2016demonstration,ballance2016high}. By adding non-Hermiticity into this system, exotic dynamics have been observed with both $\mathcal{PT}$ \cite{wang2021observation, ding2021experimental}and anti-$\mathcal{PT}$-  symmetric non-Hermitian Hamiltonian \cite{ding2022information,bian2022quantum}. It is naturally to ask if such systems can support a quantum logic gate much faster than Hermitian quantum mechanics, and if such a gate could approach the QSL in an open system. This letter address these questions by reporting the first experimental realization of the speeding-up of  single-qubit gates in trapped ion with $\mathcal{PT}$-symmetric Hamiltonian, and further test the tight bound of QSL in an open qubit system.

It is noted that the speed-up of non-Hermitian system has been confirmed in an optical cavity QED system \citep{cimmarusti2015environment} and work-ancilla two-qubit nuclear magnetic resonance system
\citep{zheng2013observation}. In the former, the speed-up resulted from the environment induced coupling change in the off-diagonal term of the Hamiltonian, while in the latter, it resulted from the post-selection of the non-Hermitian elements of a Hermitian system.  In this work, the speed-up is realized with truly controllable dissipation in a pure qubit system, enabling the capability to implement the speed-up gate operation in a scalable system having finite dissipation to the environment. Such a vision also addresses David DiVincenzo's argument how to initialize a large scale quantum system strongly coupled to the environment ``... the Hamiltonian of the system and its environment are necessarily perturbed strongly, ... but potentially much shorter than the natural relaxation times" when discussing the five requirements for the physical implementation of quantum computation \cite{divincenzo2000physical}.

First we have experimentally demonstrated such a fast evolution based on a single $^{171}$Yb$^+$ ion qubit system, by carefully controlling the dissipation strength. We show the speed is beyond the QSL defined by relative purity metric, but exactly matches with the limit predicted by Fubini-Study metric. Then, to explain the speed up induced by the non-Hermiticity, we map the $\mathcal{CPT}$ inner product to the Hilbert space by constructing a variant Bloch sphere with the $\mathcal{CPT}$ theorem \cite{bender2007faster}. This allows us to find out the optimal path -- the geodesic line in the variant Bloch sphere -- in speeding up the qubit operation.

\begin{figure}
   \includegraphics[width=1.0\linewidth]{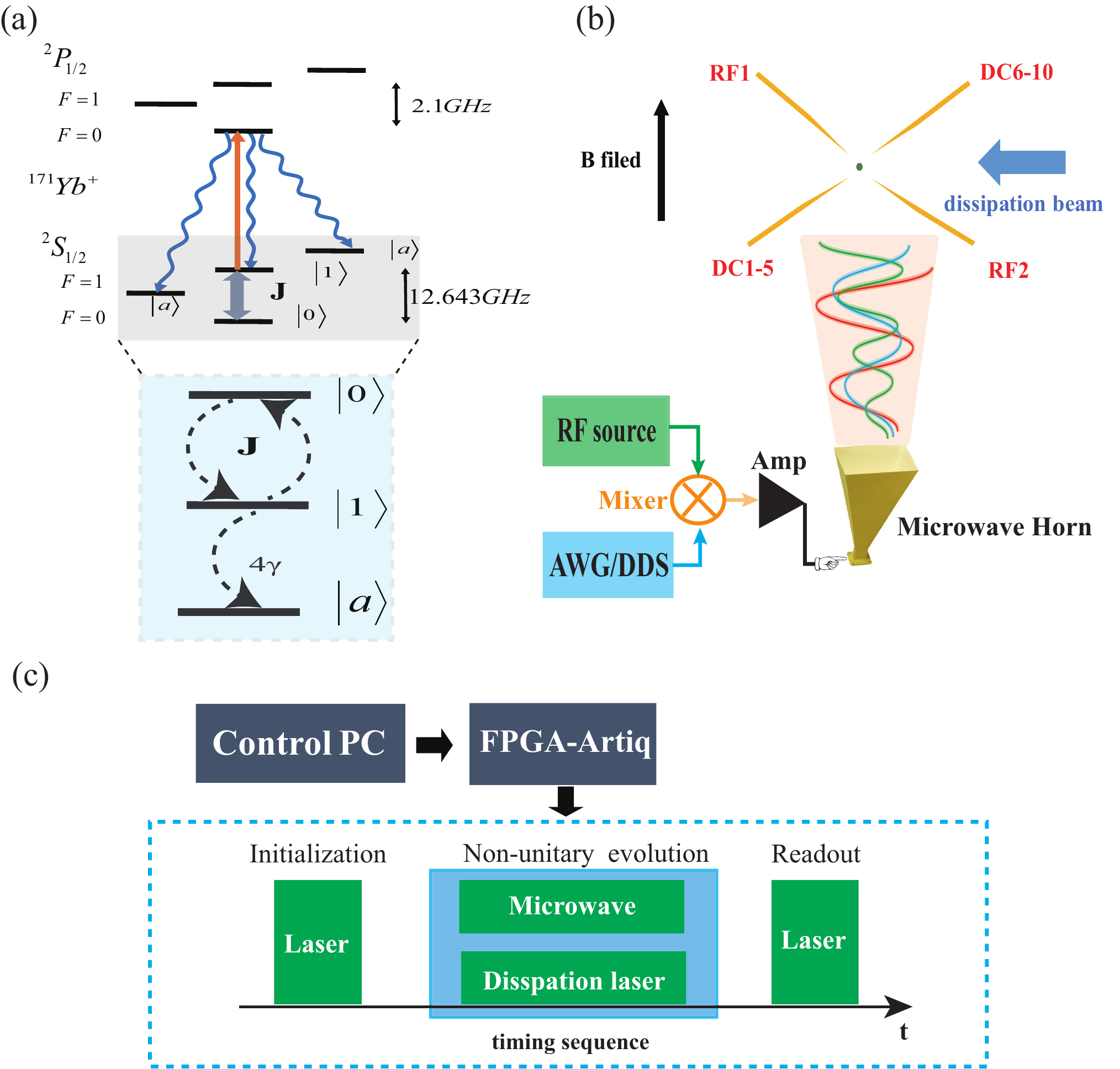}
    \caption{The experimental system of realizing passive non-Hermitian $\mathcal{PT}$-symmetric Hamiltonian. (a) The related energy levels of the $^{171}$Yb$^{+}$ ion . The involved five levels $|F=0, m_F =0\rangle$ and $|F=1, m_F =0, \pm1\rangle$ in electronic ground state $^2 S_{1/2}$, and $|F=0, m_F =0\rangle$ in electronic excited state $^2 P_{1/2}$ can be simplified to a three-level non-Hermitian system with $|0\rangle = |^2 S_{1/2}, F=0, m_F =0\rangle$, $|1\rangle = |^2 S_{1/2}, F=1, m_F =0\rangle$ and $|a\rangle = |^2 S_{1/2}, F=1, m_F = \pm1\rangle$, where population on $|1\rangle$ dissipate to $|a\rangle$ with a rate of $4\gamma$. (b) Schematic diagram of the experimental setup. The magnetic (B) field, the dissipation beam and the microwave (MW) are illustrated. (c) Timing control in the experiment. }
\label{setup}
\end{figure}

\noindent\textit{\textbf{Experiments}}\quad In order to experimentally verify this speed-up evolution enabled by the $\mathcal{PT}$-symmetric quantum theory, we constructed a passive $\mathcal{PT}$-symmetric system \cite{ding2021experimental} in a trapped $^{171}$Yb$^{+}$ ion system. The related energy levels of the $^{171}$Yb$^{+}$ ion are shown in the
Fig.\ref{setup}(a).  Details are shown in supplementary materials S1. A
369.5 nm laser beam with adjustable intensity, which contains only $\pi$ polarization component in
the vertical direction, was employed as a dissipation beam to excite the ion from $|1\rangle$ to $|F
= 0, m_{F} = 0\rangle$ in $^{2}P_{1/2}$ electronic excited state. The ion will spontaneously decay
to either one of the three states ($|F = 1, m_{F} = 0,\pm1\rangle$) of $^{2}S_{1/2}$ with the same
probability. This configuration can be simplified to a two-level open system $|0\rangle$ and
$|1\rangle$, with $|a\rangle = |F = 1, m_{F}\pm1\rangle$ taken as the environment. In the presence
of coupling, the two-level open system can be described by an effective non-Hermitian Hamiltonian
($\hbar$=1):
\begin{equation}
{H_{eff}} = J\left( {\left|  0  \right\rangle \left\langle  1  \right| + \left|  1  \right\rangle \left\langle  0  \right|} \right) - 2i\gamma \left|  1  \right\rangle \left\langle1  \right|,
 \label{Heff}
\end{equation}
where $J$ is coupling rate and $\gamma$ is dissipation rate. It can be mapped to the non-Hermitian
$\mathcal{PT}$-symmetric Hamiltonian in Eq.(\ref{Hpt}) by adding an identity matrix:
\begin{equation}
H_{PT}=H_{eff}+i\gamma \pmb{I},
\label{hpt2}
\end{equation}
which describes a balanced loss and gain two-level system. More details can be found in supplementary materials S2.

\noindent\textit{\textbf{Results}}\quad In our system, the coupling rate $J$ and dissipation rate $\gamma$ can be adjusted
by tuning the intensity of microwave and dissipation beam, respectively. First, we prepared the system in the initial state $|1\rangle$, then we detected the population in
state $|1\rangle$ after a certain time of non-unitary evolution, which is given by:
\begin{equation}
{\rho _{11 }}\left( \tau_f \right)={{\rm{e}}^{ - 2\gamma \tau_f}}{\left[ {\cos \left( {\chi \tau_f} \right) - \frac{{\gamma \sin \left( {\chi \tau_f} \right)}}{\chi }} \right]^2},
\label{r11}
\end{equation}
where $\chi =\sqrt {J^2 - \gamma ^2}$. See supplementary materials S3 for detailed derivation. The coupling rate $J$ can be determined from the measurement of Rabi oscillation. For a fixed $J$, the dissipation rate $\gamma$ can be obtained by fitting the population curve according to Eq.(\ref{r11}). The state flipping time $\tau_f$ from $|1\rangle$ to $|0\rangle$ under non-Hermitian evolution is taken as the time when the population of state $|1\rangle$ vanishes to 0, and can be expressed by:
\begin{equation}
\tau_f = \frac{[{\pi - 2\arcsin \left( {\frac{\gamma }{J}} \right)}]\hbar}{{2J\sqrt {1 - {{\left( {\frac{\gamma }{J}} \right)}^2}} }}.
\label{t}
\end{equation}
It gets smaller with larger $\gamma/J$ and is called \emph{wormholelike effect} in Ref.{\cite{bender2007faster}}. More details can be found in supplementary materials S4.

\begin{figure}
\centering
\includegraphics[width=1.0\linewidth]{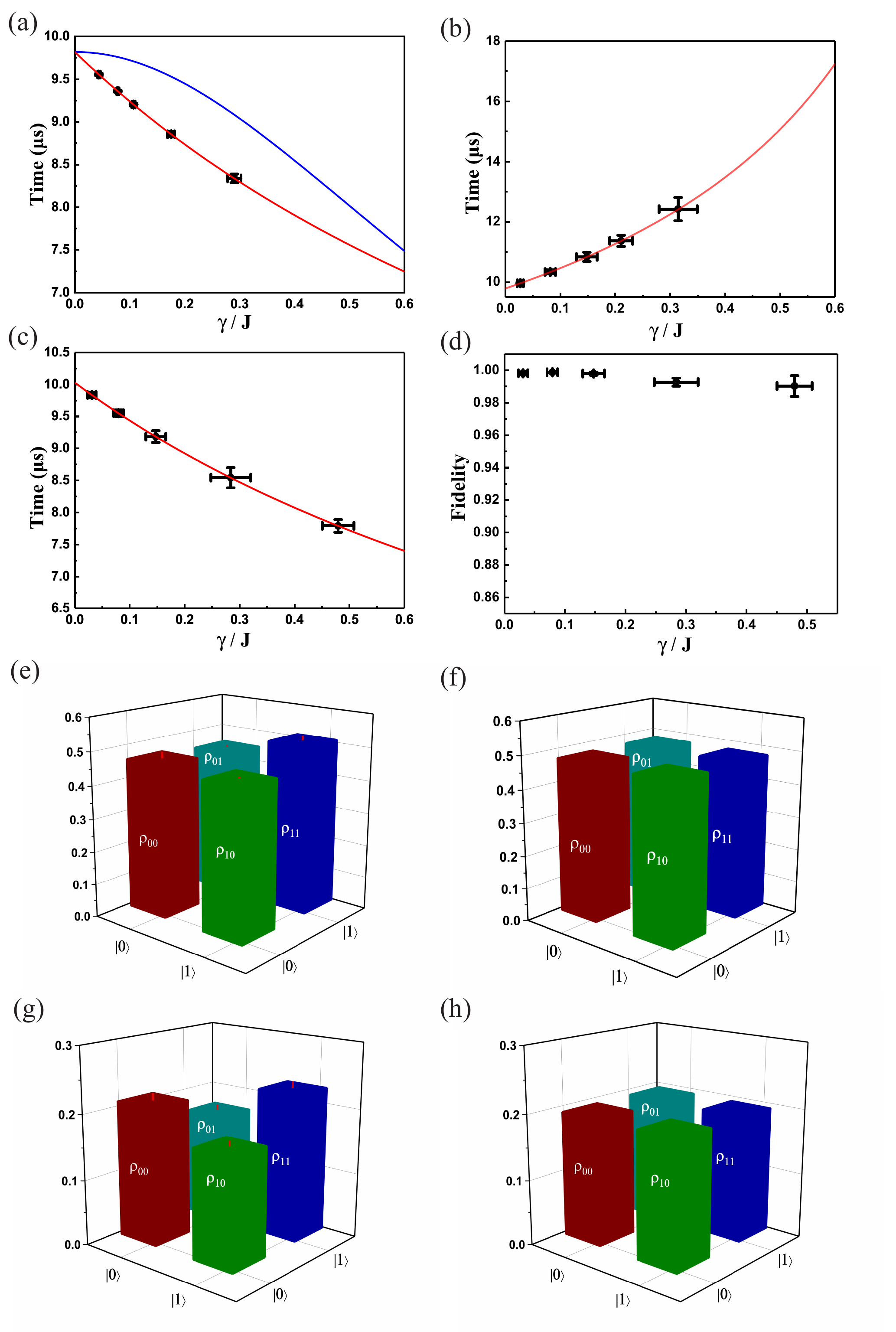}
\caption{The evolution of single trapped ion-qubit gate. The evolution time (a) from $|1\rangle$ to $|0\rangle$, (b) from $|0\rangle$ to $|1\rangle$, (c) from $|1\rangle$ to $|\psi_{s}\rangle$ under different dissipation
intensities $\gamma/J$. The black circles and error bars are estimated from experiment results. The red line in (a, c) and (b) correspond to the theoretical curve described by Eq.\ref{t} and Eq.\ref{t2}, respectively. The blue line in (a) corresponds to the proposed QSL in a open system \cite{del2013quantum}. (d) The process fidelity obtained from measurements of quantum state tomography for five different $\gamma/J$ in (c) . The experimentally and numerically reconstructed density matrix at $\tau_s= 4.92\mu s$, $\gamma/J=0.03$ corresponding to the first point in (c) are illustrated in (e) and (f), respectively. The experimentally and numerically reconstructed density matrix at $\tau_s= 3.90\mu s, \gamma/J=0.48$ corresponding to the fifth point in (c) are illustrated in (g) and (h), respectively. Each density matrix is detected by five rounds and each round preserve average data of ten measurements. }
\label{fast}
\end{figure}

We measured this flipping time $\tau_f$ under different dissipation intensity $\gamma/J$ and found good agreement with theoretical predictions as shown in Fig.\ref{fast}(a). When $\gamma/J = 0$, $\tau_f=\pi \hbar /2J$, corresponding to the Rabi flipping time in an unitary evolution. We noted it equaled both Mandelstan-Tamm bound and Margolus-Levitin bound, according with the prediction of QSL in the Hermitian system. As the dissipation strength was further increased ($0<\gamma <J$), it became smaller and smaller. This agrees with the generalized Margolus-Levitin bound for the QSL based on Bures metric in an open system \cite{deffner2013quantum}. Compared with the QSL $\tau = \frac{\pi \hbar}{2\sqrt{J^2 + 2\gamma^2}}$ based on the metric of relative purity \cite{del2013quantum}, it is significantly smaller, thus representing a tighter bound for QSL.

We also investigated the dependence of the flipping time $\tau_{f}^{\prime}$ from $|0\rangle$ to $|1\rangle$ state on the $\gamma/J$, the results is shown in Fig.\ref{fast}(b). Theoretically, the flipping time $\tau^{\prime}_{f}$ from $|0\rangle$ to $|1\rangle$ follows
\begin{equation}
 \tau^{\prime}_{f}= \frac{{\pi + 2\arcsin \left[ {\frac{\gamma }{J}} \right]}}{{2J\sqrt {1 - {{\left( {\frac{\gamma }{J}} \right)}^2}} }},
\label{t2}
\end{equation}
which gets larger with larger $\gamma/J$. This is evident in Fig.\ref{fast}(b).

Furthermore, we performed another qubit operation, i.e. rotating the qubit from
initial state $|1\rangle$ to a normalized superposition state $|\psi_{s}\rangle =
\frac{1}{\sqrt{2}}(|1\rangle-i|0\rangle)$. By measuring the time-dependent population in state
$|1\rangle$ and fitting it according to Eq.(\ref{r11}), the evolution time $\tau_{s}$ of
$|1\rangle\to|\psi_{s}\rangle$ was acquired, as shown in Fig.\ref{fast}(c). Indeed, it is smaller than $\tau$ and become smaller and smaller as $\gamma$ increases, exactly as what
Eq.(\ref{r11}) has predicted.

The fidelity of qubit was checked for each operation \cite{wang2008alternative}. As seen From Fig.\ref{fast}(d), all of the measured fidelities are nearly unity during the operation, certifying the coherent evolution process. This was done by reconstructing the density matrix $\rho_{a}=|\psi_{a}\rangle\langle\psi_{a}|$ based on the quantum state tomography, as shown in Fig.\ref{fast}(e-h). Here $|\psi_{a}\rangle$ is defined as $|\psi_{a}\rangle=e^{-iH_{eff}\tau_{s}/\hbar}|1\rangle$. In order to perform the tomography, we prepared the state to $|\psi_{a}\rangle$ through precisely controlling the duration of both dissipated laser and microwave. In Fig.\ref{fast}(e), the measurements of four matrix elements for $\gamma/J=0.03$ and $t_{s}= 4.92\mu s$ are plotted against theoretical values plotted in
Fig.\ref{fast}(f). Similarly, the comparison between measurements and theoretical calculations is
illustrated in Fig.\ref{fast}(g-h) for $\gamma/J=0.48$ and $t_{s}= 3.90\mu s$. Evidently,
the measurements agreed with theoretical calculations pretty well.

\begin{figure}
\centering
\includegraphics[width=1.0\linewidth]{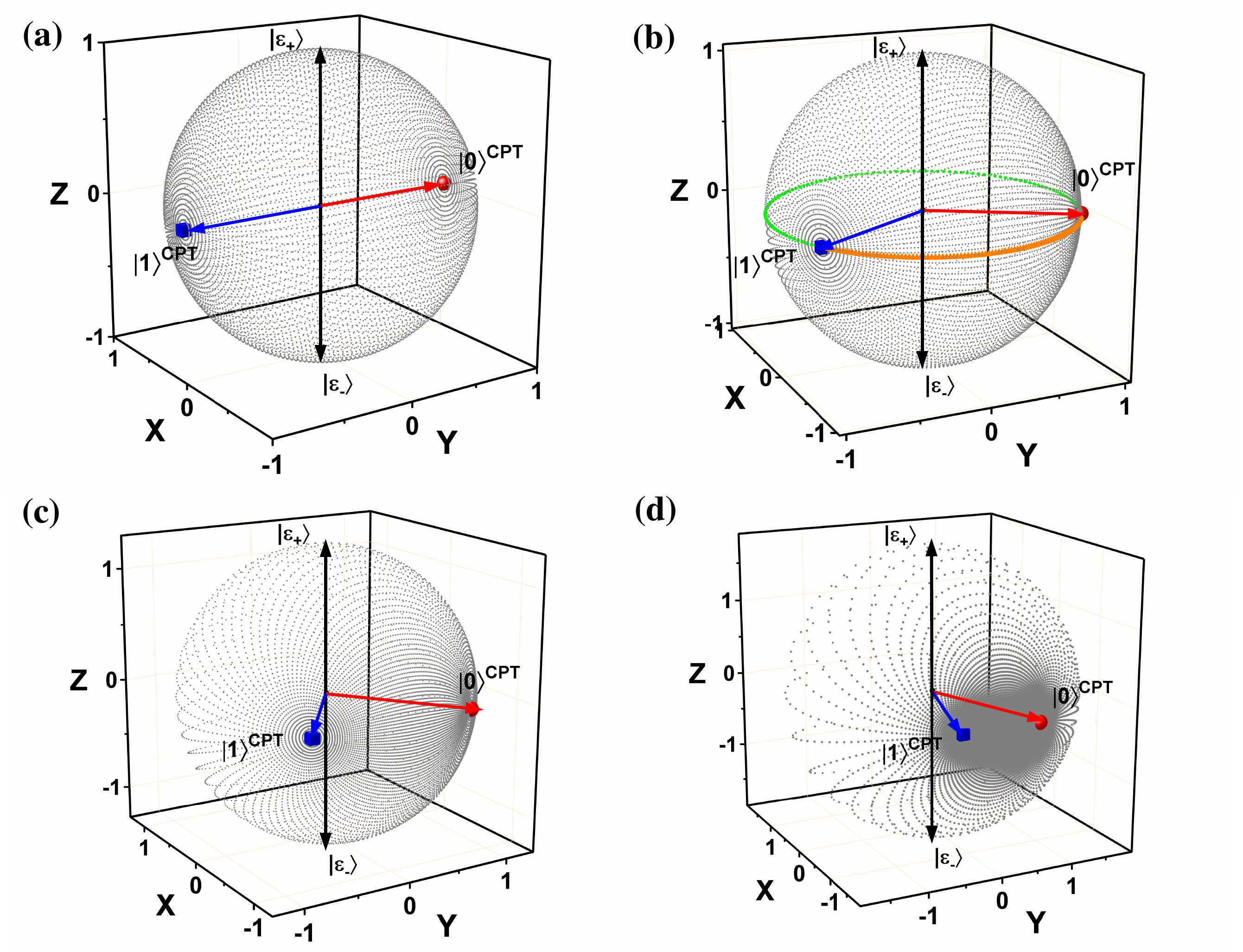}
\caption{Non-Hermitian $\mathcal{PT}$ symmetry Bloch sphere mapped from Hermitian Bloch sphere. The red sphere (blue cube) represents $|1\rangle^{\mathcal{CPT}}$($|0\rangle^{\mathcal{CPT}}$) mapped from state $|1\rangle$($|0\rangle$) described in Hermitian space. We scan $\theta$ from 0 to $\pi/2$ and $\phi$ from 0 to $2\pi$, taking 10100 points on average in Hermitian Bloch sphere. The $\gamma$ in four pictures are (a) $ 0J$, (b) $ 0.4793J$, (c) $ 0.8J$ and (d) $0.96J$, respectively. The orange and green curves in (b) represent the trajectories of the evolution from $|1\rangle$ to $|0\rangle$ and from $|0\rangle$ to $|1\rangle$, respectively. }
\label{sphere}
\end{figure}

\noindent\textit{\textbf{QSL with the non-Hermitian Bloch sphere}}\quad
For a general $\mathcal{PT}$-symmetric Hamiltonian:
\begin{equation}
H_{PT}=\left(\begin{array}{cc}
i\gamma & J e^{i\phi_J} \\
J e^{-i\phi_J} & -i\gamma
\end{array}\right),
\label{Hpt}
\end{equation}
the minimum evolution time of a qubit from $|0\rangle$ to $|1\rangle$ keeps the form of Eq. (\ref{t}). We interpreted this from the geometric point of view. In a Hermitian system, the evolving time from an initial state to a final state is interpreted using the Fubini-Study metric in the Hilbert space \cite{anandan1990geometry,brody2003elementary}. In a non-Hermitian system, treating this problem with geometric analysis has been theoretically discussed \cite{gunther2008p,giri2008lower}. The idea that the QSL is determined by the geodesic line in the transformed Hilbert space metric using $\mathcal{CPT}$ theorem has been proposed \cite{giri2008lower}. Hence, we developed a method to visualize the deformation of the Fubini-Study metric of the Hilbert space by constructing the $\mathcal{CPT}$-symmetric non-Hermitian Bloch sphere and state vectors related to Eq.(\ref{Hpt}). We noticed that a special case of Eq.(\ref{Hpt}) with $\phi_J = 0$ had been visualized using the similar approach in Ref.\cite{gunther2008p}.

The linear operator $\mathcal{C}$, which satisfies $\mathcal{C}^2=1$ and commutes with both the Hamiltonian $H$
and the operator $\mathcal{PT}$, has been defined to construct a new form of inner product\cite{bender2002complex,bender2005introduction}. This inner product is defined as $\langle a | b \rangle^{\mathcal{CPT}}=(\mathcal{C P T} a) \cdot b$, where the complex conjugate is replaced by the $\mathcal{CPT}$-conjugate
\cite{bender2004scalar}. For a certain  $|\psi\rangle$, the value of the new inner product can be
written as the form of Dirac inner product:
\begin{equation}
\langle\psi|\psi\rangle^{\mathcal{CPT}}=\langle\psi| \mathcal{P}^{T}\mathcal{C}^{T}|\psi\rangle
\label{cpt}
\end{equation}

By using the $\mathcal{CPT}$ conjugate, the eigenstates $|\varepsilon_{\pm}\rangle$ of $H_{PT}$ in Eq.(\ref{Hpt}) can be regarded as a set of complete orthogonal basis, and the structure of the new Hilbert space depends on the form of the operator $\mathcal{C}$. For the more general cases, where the off-diagonal terms in Eq.(\ref{Hpt}) has a specific phase $\phi_J$, the operator $\mathcal{C}$ is expressed by:
\begin{equation}
\mathcal{C}=\frac{1}{\sqrt{1-(\frac{\gamma}{J})^2}}\left(\begin{array}{cc}
i e^{-i\phi_J} \frac{\gamma}{J} & 1 \\
1 & -i e^{i\phi_J} \frac{\gamma}{J}
\end{array}\right)
\label{C}
\end{equation}
Here, the axis of qubit rotation is generalized to arbitrary axis in the X-Y plane. When $\phi_J=0$, the axis of qubit rotation become the X axis. We find that in this more general case, the evolution along the geodesic line is the fastest among all possible path.

Based on Eq.(\ref{cpt}) and Eq.(\ref{C}), we established a mapping between a Hermitian Bloch
sphere and a $\mathcal{PT}$-symmetric non-Hermitian Bloch sphere, which can be described as:
\begin{equation}
f:|\psi\rangle\rightarrow|\psi\rangle^{\mathcal{CPT}}
\end{equation}
The state $|\psi\rangle=\cos \frac{\theta}{2}|0\rangle+e^{i \phi} \sin \frac{\theta}{2}|1\rangle$
is well described in a normal Hermitian Bloch sphere with the orthogonal basis $|0\rangle$ and
$|1\rangle$. In $\mathcal{CPT}$-conjugate Hilbert space, the eigenstates $|\varepsilon_+\rangle$ and $|\varepsilon_-\rangle$ of $\mathcal{P T}$-symmetric
Hamiltonian are orthogonal. Using
$|\varepsilon_+\rangle$ and $|\varepsilon_-\rangle$ as the basis, $|\psi\rangle^\mathcal{CPT}$ can
be written as:
\begin{equation}
|\psi\rangle^{\mathcal{CPT}}=R \cos \frac{\Theta}{2}\left|\varepsilon_{+}\right\rangle+R \sin \frac{\Theta}{2} e^{i \Phi}|\varepsilon_{-}\rangle,
\end{equation}
where $R$, $\Theta$ and $\Phi$ are functions of $\theta$, $\phi$ and $\phi_J$ :
\begin{equation}
\begin{array}{l}
R=R(\theta, \phi,\phi_J), \Theta=\Theta(\theta, \phi,\phi_J),  \Phi=\Phi(\theta, \phi,\phi_J)
\end{array}
\end{equation}
More details can be found in supplementary materials S5.

By fixing the value of R to $\sqrt{\langle1|1\rangle^\mathcal{CPT}}$, a $\mathcal{PT}$-symmetric
non-Hermitian Bloch sphere can be constructed, where $|1\rangle^\mathcal{CPT}$ evolves under unitary evolution. We note the $\mathcal{P T}$-symmetric
non-Hermitian Bloch sphere can be regarded as the promotion of Hermitian Bloch sphere. This is illustrated in Fig.\ref{sphere}(a), when $\gamma=0$, the Hamiltonian in Eq.(\ref{Heff}) becomes Hermitian, and the non-Hermitian Bloch sphere turns into a Hermitian Bloch sphere. As $\gamma$ gradually increases, due to the effect of $C$ operator, the non-Hermitian Bloch sphere get more and more distorted. More importantly, the
geodesic "distance" between $|0\rangle$ and $|1\rangle$ become shorter and shorter with increasing
$\gamma/J$, as shown in Fig.\ref{sphere}(b-d). As $\gamma/J$ approaches 1, i.e. the exceptional point, the ``distance" become infinitesimally small. Consequently, the qubit evolution time will become smaller and smaller with increasing dissipation $\gamma$ for fixed coupling $J$. This has been experimentally verified as mentioned above. It can be understood since the evolution makes a longer trip along the geodesity on the Bloch sphere as illustrated in Fig.\ref{sphere}(b). This can also be verified by the geometrical relation that the whole period for a round trip from state $|1\rangle$ to $|0\rangle$, and back to $|1\rangle$, should not be smaller than two fold of the Margolus-Levitin bound \cite{giri2008lower}. It is in analogy to the distortion of space-time by the gravitation field in Einstein's general theory of relativity.

\noindent\textit{\textbf{Conclusion}}\quad We have experimentally verified a faster-than-Hermitian single qubit gate using a trapped-ion qubit with a constructed $\mathcal{PT}$-symmetric Hamiltonian. The evolution time gets shorter with increasing dissipation
strength as long as it is smaller than the coupling strength, i.e. the system remains in the
$\mathcal{PT}$-symmetric phase. Meanwhile, the fidelity of the quantum state evolution remains
nearly unity despite of increasing dissipation strength. The evolution speed indicates a tighter QSL than the QSL derived with a relative purity metric in open quantum system, while it can be well explained using Fubini-Study metric for non-Hermitian Bloch sphere. Such a speedup holds great promises to open new vistas for quantum computing in the presence of the interaction with the environment.

\noindent\textit{\textbf{Acknowledgements}}\quad This work is supported by the Key-Area Research and Development Program of Guang Dong Province under Grant No.2019B030330001, the National Natural Science Foundation of China under Grant No.11774436, No.11974434 and No. 12074439, Natural Science Foundation of Guangdong Province under Grant 2020A1515011159, Science and Technology Program of Guangzhou, China 202102080380, the Central -leading-local Scientific and Technological Development Foundation 2021Szvup172. Le Luo acknowledges the support from Guangdong Province Youth Talent Program under Grant No.2017GC010656.

\clearpage
\widetext
\begin{center}
\textbf{\large Supplementary Materials for: Realizing quantum speed limit in open system with a PT-symmetric trapped-ion qubit}
\end{center}
\setcounter{equation}{0}
\setcounter{figure}{0}
\setcounter{table}{0}
\setcounter{page}{1}
\setcounter{section}{0}
\makeatletter
\renewcommand{\theequation}{S\arabic{equation}}
\renewcommand{\thefigure}{S\arabic{figure}}
\renewcommand{\bibnumfmt}[1]{[S#1]}
\renewcommand{\citenumfont}[1]{S#1}
\renewcommand{\thesection}{S\arabic{section}}

\section{Experimental setup}

Our experimental setup has been described in detail elsewhere \cite{liu2021minimization1}. A single
$^{171}$Yb$^+$ ion is confined and laser cooled in a linear Paul trap consisting of four
gold-plated ceramic blade electrodes. The schematics of the trap and related energy levels of the
ion are shown in Fig.1(a) and (b), respectively. The radio frequency (RF) signal and DC
voltages are applied to two RF electrodes (RF1 and RF2) and two DC electrodes, respectively. Each
DC electrode is divided into five segments. We define trap axis as $\textbf{X}$ axis, other two
axes perpendicular to the trap axis as $\textbf{Y}$ and $\textbf{Z}$ axis. The trap has an axial
trap frequency of $\nu_x=2\pi \times 0.744$ MHz and two radial trap frequency of ($\nu_y,
\nu_z$)=($2\pi \times 1.382$ MHz, $2\pi \times 1.655$ MHz).

In the system, a mixed microwave signal which consists of a 12.611580 GHz signal from standard RF
source (Rohde and Schwarz, SMA 100B) and a 31.25 MHz signal from an arbitrary waveform generator
(AWG, Spectrum Instrumentation) drives the qubit rotation. The relative phase of the microwave is
precisely controlled by the input functions and parameters on AWG. A 1 GHz signal from standard RF
source is used as the AWG reference. A pair of Helmholtz coils create a magnetic field around 6
Gauss along vertical $\textbf{Z}$ axis, which not only shifts the degeneracy of the three magnetic
sublevels, but also prevents the ion from getting pumped into a coherent dark state.

The single trapped ion can be initialized to either $|0\rangle=|F = 0, m_{F} = 0\rangle$ or
$|1\rangle=|F = 0, m_{F} = 1\rangle$ in $^{2}S_{1/2}$ electronic ground state. A qubit is constructed by
driving a microwave transition with the mixed signal. The excitation
from $|0\rangle=|F = 0, m_{F} = 0\rangle$ in $^{2}S_{1/2}$ to $|F = 1, m_{F} = 0\rangle$ in
$^{2}P_{1/2}$ electronic excited state is driven by a 369.5 nm dissipation laser beam, which
contains only $\pi$ polarization component in the vertical direction, leading to spontaneous decay
to $|F = 1, m_{F} = 0, \pm 1\rangle$ three magnetic sublevels in $^{2}S_{1/2}$ electronic ground
state with equal probability. The decay to $|F = 1, m_{F} = \pm 1\rangle$ can be taken as
equivalent loss of the two-level qubit system, resulting in a non-unitary evolution. Thus, we
construct a passive $\mathcal{P T}$-symmetric system \cite{ding2021experimental1} in a trapped
$^{171}$Yb$^+$  ion system by simultaneously driving the microwave and optical transition to the
ion.

For the timing control, the switching of the cooling laser, pumping laser,
dissipation beam, and detection beam are all controlled by switching acoustic-optic modulators (AOM,
Brimose TEM-200-50-369) with RF switches (Mini-Circuits, ZASWA-2-50DR+), to which TTL signals are
fed from ARTIQ (Advanced Real-Time Infrastructure for Quantum physics) device (M-Labs, Sinara
Kasli). The synchronization of the microwave and dissipation beam is precisely controlled. The
intensity of the dissipation beam can be tuned by adjusting the RF power applied on the AOM.

The population of state
$|1\rangle$ or $|0\rangle$ was measured by the standard fluorescence counting rate threshold method. A 369.5 nm detection beam which contains $\sigma_{\pm}$ and $\pi$
polarization, was propagated along X axis. Note that $|1\rangle$ cannot be
detected directly because the detection beam simultaneously pumps all three Zeeman levels ($|F = 1,
m_{F} = 0,\pm1\rangle$), leading to indistinguishable population information. The population information in $^{2}S_{1/2}$ $|F = 0, m_{F} = 0\rangle$ can be precisely determined since total probability in all four states ($|0\rangle$, $|1\rangle$ and $|a\rangle$) equals 1. For
example, we can determine the population in state  $^{2}S_{1/2}$ $|F = 1, m_{F} = 0\rangle$ by
applying an extra Rabi $\pi$ flip to exchange the population of $|0\rangle$ and $|1\rangle$ prior
to the detection.

\section{The properties of  a $\mathcal{PT}$ -symmetric in non-Hermitian system}

\begin{figure}
\begin{center}
\setlength{\abovecaptionskip}{0.cm}
\setlength{\belowcaptionskip}{-0.cm}
\includegraphics[width=1.0\linewidth]{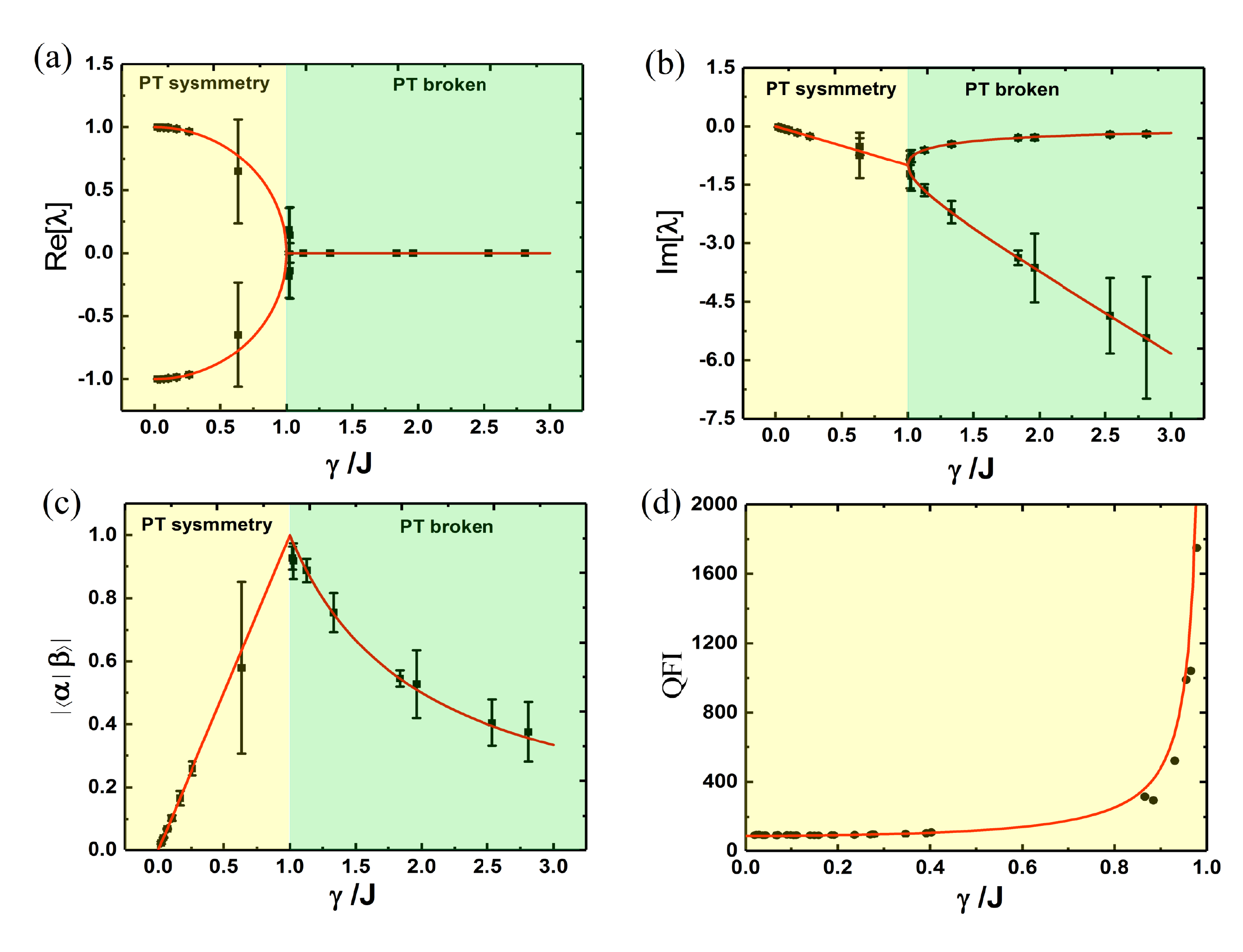}
\caption{The real part (a) and the imaginary part (b) of the eigenvalues. In both (a) and (b), the
rectangles represent the experimental data, while the red line represents the theoretical solution.
In (b), the imaginary part in the $\mathcal{PT}$-symmetric phase is negative and decreases with
$-i\gamma$ due to a passive $\mathcal{PT}$-symmetric system. It separates into a slow (upper
branch) and a fast(lower branch) decay mode in the $\mathcal{PT}$-broken phase. (c) The overlap
between the two eigenstates $|\alpha \rangle$ and $|\beta \rangle$  in both $\mathcal{PT}$-symmetric
and $\mathcal{PT}$-broken phases, satisfying $|\langle \alpha | \beta \rangle | = \min(\gamma
/J, J/\gamma)$ (red line). (d) The theoretical quantum Fisher information (QFI) of damping rate
\citep{xie2019enhancing} $F=\frac{2}{J^2 - \gamma ^2}$  (red line) for $J > \gamma$, compared with
the experimental measurement (black circles).} \label{eigen}
\end{center}
\end{figure}

We investigated the $\mathcal{PT}$ symmetry breaking transition. In Fig.\ref{eigen}(a-b), the
evolution of the eigenvalues in a passive $\mathcal{PT}$ system is shown with $\gamma$ for a constant $J$. At
$\gamma=0$, the system is a Hermitian system and has two real eigenvalues. As it increases from 0
to $J$, the imaginary part of the eigenvalue Im[$\lambda$] appears and decreases linearly with
$-i\gamma$ from 0, while two real parts Re[$\lambda$] follow the upper and lower halves of the
circle $\sqrt{J^2-\gamma ^2}$ and gradually converge. At $\gamma=J$, Re[$\lambda$]=0 and
Im[$\lambda$]=$-\gamma$, this is so-called exceptional point (EP). When it further increases, the
real part stays 0, whereas the imaginary part separates into two modes, one is slow decay mode
(upper branch) and follows the hyperbolic line according to $-\gamma+\sqrt{{\gamma^2}-J^2}$, while
the other is fast decay mode (lower branch) and follows the hyperbolic line according to
$-\gamma-\sqrt{{\gamma^2}-J^2}$. The region of $0\le \gamma < J$ is defined as
$\mathcal{PT}$-symmetric phase, whereas the region of $\gamma > J$ is defined as
$\mathcal{PT}$-broken phase. The overlap between $|\alpha \rangle$ and $|\beta \rangle$ two
eigenstates in both $\mathcal{PT}$-symmetric and $\mathcal{PT}$-broken region, which satisfies
$|\langle \alpha | \beta \rangle | = \min(\gamma /J, J/\gamma)$, is shown in Fig.\ref{eigen}(c).
Near the EP point, the two eigenstates of Hamiltonian coalesce into one.

\begin{figure}
\setlength{\abovecaptionskip}{0.cm}
\setlength{\belowcaptionskip}{-0.cm}
\includegraphics[width=1.0\linewidth]{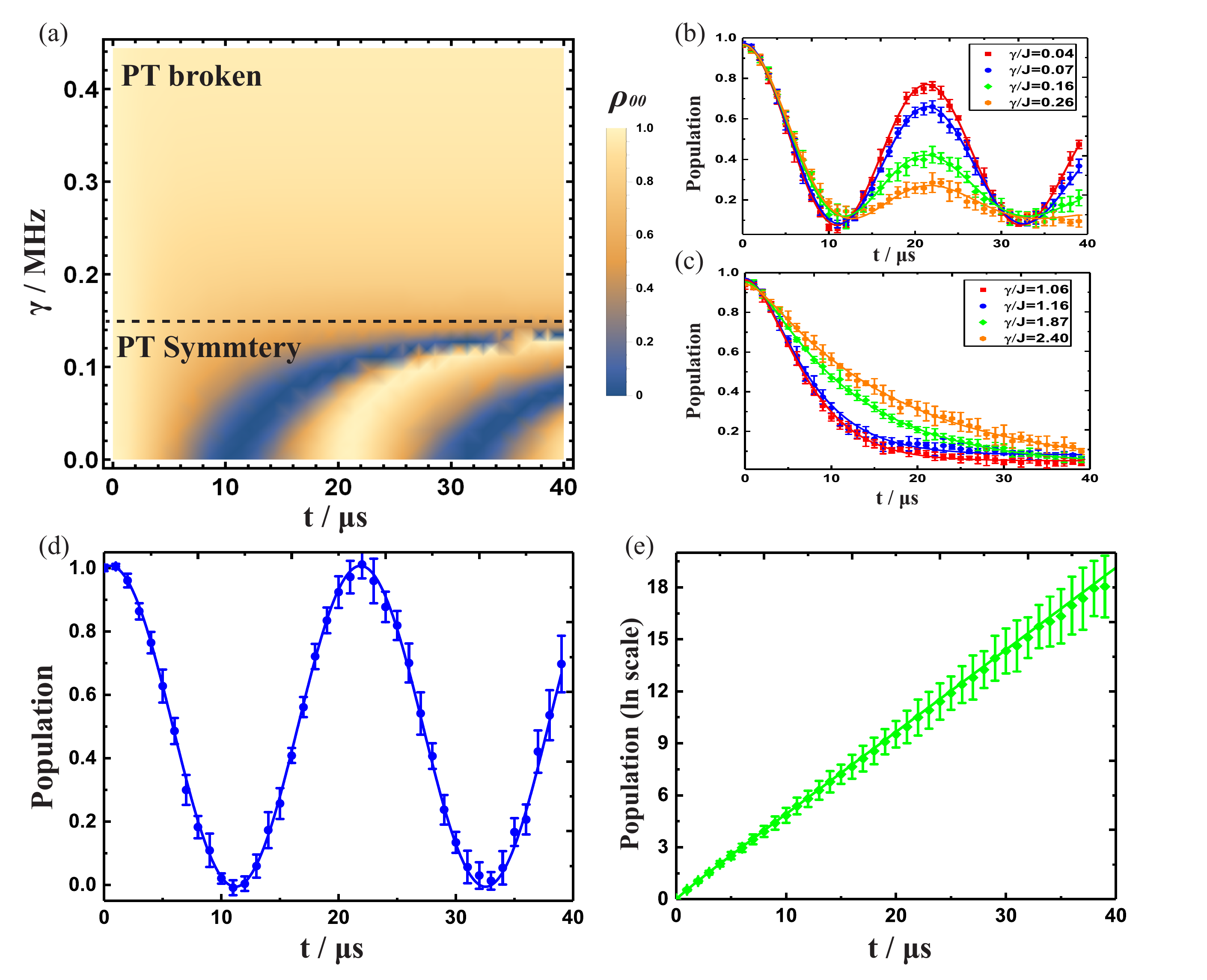}
\caption{(a) Color maps of the population on $|0\rangle$ versus $t$ for various dissipation
strengths $\gamma$. In our system, the coupling between $|0\rangle$ and $|1\rangle$ is kept at $J =
0.148$ MHz. The evolution of the population in $|0\rangle$ with time for different $\gamma /
J$ in both dissipative $\mathcal{PT}$-symmetric phase (b) and $\mathcal{PT}$-broken phase (c),
illustrating the evolution dynamics between $|0\rangle$ and $|1 \rangle$. In (b), red squares,
blue circles, green diamonds and yellow hexagons correspond to $\gamma /J = 0.04$, 0.07, 0.16 and
0.26, respectively. In (c), red squares, blue circles, green diamonds and yellow hexagons
correspond to $\gamma /J = 1.06$, 1.16, 1.87 and 2.40, respectively. (d) The evolution of the
population in $|0 \rangle$ with time for $\gamma / J = 0.07$ in dissipative
$\mathcal{PT}$-symmetric phase is mapped to $\mathcal{PT}$-symmetric system with balanced gain and
loss, since both of the systems share the same topological features. (e) The evolution of $|0\rangle$ state with time for $\gamma / J = 1.87$ in passive $\mathcal{PT}$-broken phase is mapped to $\mathcal{PT}$-symmetric system with balanced gain and loss.}
\label{NH}
\end{figure}

In addition, the eigenstates of the non-Hermitian Hamiltonian $H_{eff}$ can be used for measuring
$\gamma$. The explicit formula of the quantum Fisher information (QFI) for the two-dimensional
density matrices is written as \cite{zhong2013fisher}:
\begin{equation}
\mathcal{F}_{\gamma }=\text{Tr}\left(\partial_{\gamma}\rho \right)^2+\text{Tr}\left(\rho \partial_{\gamma} \rho \right)^2/{\det\rho}.
\label{fisher}
\end{equation}
It can be used to obtain the QFI of damping rate $F=\frac{2}{J^2 - \gamma ^2}$ for $J
>\gamma$, as is shown in Fig. \ref{eigen}(d). As $\gamma$ approaches $J$, the magnitude of QFI
goes to infinity. We show here the location of the EP can be precisely determined by utilizing this feature of QFI. One can also employ this feature to obtain a very good precision of damping rate $\gamma$ \cite{xie2019enhancing} by tuning the coupling
strength $J$ to EP. It is worth to notice, the enhanced sensitivity near the EP bears similarities
to weak value amplification\cite{jordan2014technical, naghiloo2019quantum}.

Then, we directly measured the density matrix elements of $\rho_{00}(t)$ by performing non-Hermitian
evolution, thus obtaining the population in $|0\rangle$ state. Color maps of the population on
$|0\rangle$ versus $t$ for various dissipation strengths $\gamma$ are shown in Fig. \ref{NH}(a). The
evolution dynamics of the population is further studied by comparing it in $\mathcal{PT}$-symmetric
phase (Fig. \ref{NH}(b)) with the one in $\mathcal{PT}$-broken phase (Fig. \ref{NH}(c)). In Fig.
\ref{NH}(b), red squares, blue circles, green diamonds and yellow hexagons correspond to $\gamma
/J = 0.04$, 0.07, 0.16 and 0.26, respectively. When $\gamma$ increases with respect to $J$, the
population shows a damping oscillation with the damping proportional to the strength of the
$\gamma$. In Fig. \ref{NH}(c), red squares, blue circles, green diamonds and yellow hexagons
correspond to $\gamma /J = 1.06$, 1.16, 1.87 and 2.40, respectively. In $\mathcal{PT}$-broken
phase, the population decays in a single exponential form and it decays slower with larger
$\gamma/J$. It is also evident, the dynamical behavior changes from damping oscillations to
exponential decay with increasing $\gamma$, when the system transits from $\mathcal{PT}$-symmetric
phase to $\mathcal{PT}$-broken phase.

Since the passive $\mathcal{PT}$-symmetric system and the $\mathcal{PT}$-symmetric system with
balanced gain and loss share the same topological features, they have one-to-one correspondence
according to Eq.(2). Thus, the investigation of the former system offer the ability to explore the dynamics
of the latter system. Consequently, the dynamical process  such as
the evolution of the population in the dissipative $\mathcal{PT}$-symmetric system can be
accurately mapped to a balanced gain and loss $\mathcal{PT}$-symmetric system through the relation
$\rho^{\mathcal{PT}}(t)=e^{2\gamma t}\rho(t)$.  The mapped population dynamics in the balanced gain
and loss $\mathcal{PT}$-symmetric system are shown in Fig.\ref{NH}(d) and (e). In Fig.
\ref{NH}(c), the evolution of population in $|0 \rangle$ state with time for $\gamma / J = 0.1$ in
dissipative $\mathcal{PT}$-symmetric phase is mapped to a $\mathcal{PT}$-symmetric system with
balanced gain and loss. Similarly, in Fig. \ref{NH}(d), the evolution of population in $|0 \rangle$
state with time for $\gamma / J = 1.87$ in dissipative $\mathcal{PT}$-broken phase is mapped. It
can be clearly seen, the population dynamics in $|0 \rangle$ for $\gamma / J = 0.07$ retains a
Rabi-oscillation-like profile after mapping. However, the population dynamics in $|0\rangle$ for
$\gamma / J = 1.87$ displays an exponential increase with time.

\section{Lindblad evolution of the three-level system}
In the main text, we focused only on the dynamics in a qubit subsystem that is governed by
$H_{eff}$. Here, we looked at the dynamics of the entire three-level
system(Fig.1(a)), which can be described by a Lindblad master equation ($\hbar= 1$):
\begin{equation}
\frac{{d\rho }}{{dt}} =  - i\left[ {{H_C},\rho } \right] + \sum\limits_{k = 0, 1 } {{\gamma _k}\left( {{L_k}\rho {L_k}^\dag  - \frac{1}{2}\left\{ {{L_k}^\dag {L_k},\rho } \right\}} \right)},
\label{S1}
\end{equation}
where $\rho \left( t \right)$is a $3\times3$ density matrix,$\frac{{d\rho }}{{dt}}$ is the time
derivative, ${H_C} = J\left( {\left|  1\right\rangle \left\langle 0 \right| + \left| 0 \right\rangle
\left\langle 1 \right|} \right)$ is a coupling Hamiltonian in the rotating frame. The Lindblad
dissipation operators ${L_ 1 }{\rm{ = }}\left| a \right\rangle \left\langle  1 \right|$ account for
the energy decay from level $|1\rangle$  to $|a\rangle$. The dagger represents the Hermitian
conjugate. Eq.(\ref{S1}) leads to the following closed set of equations for the dynamics of the two
levels system ($|1\rangle$ and $|0\rangle$) are given by:

\begin{equation}
\frac{{d\rho }}{{dt}} = \left( {\begin{array}{*{20}{c}}
{iJ\left( {{\rho _{01}}\left( t \right) - {\rho _{10}}\left( t \right)} \right)}&{i\left( {J{\rho _{00}}\left( t \right) + 2i\gamma {\rho _{01}}\left( t \right) - J{\rho _{11}}\left( t \right)} \right)}\\
{ - 2\gamma {\rho _{10}}\left( t \right) - iJ\left( {{\rho _{00}}\left( t \right) - {\rho _{11}}\left( t \right)} \right)}&{ - iJ\left( {{\rho _{01}}\left( t \right) - {\rho _{10}}\left( t \right)} \right) - 4\gamma {\rho _{11}}\left( t \right)}
\end{array}} \right)
\label{S2}
\end{equation}

Since the drive only acts on the manifold of two spin states, the dynamics of the auxiliary state
are decoupled from the spin manifold. For a given initial condition, we can solve Eq.(\ref{S2}) to
obtain the evolution of any observable. In the experiment, the system was initialized in the
$|1\rangle$ state. The evolutions of spin levels in the $\mathcal{P
T}$-symmetric are given by:
\begin{equation}
\begin{aligned}
{\rho _{00 }}\left( t \right) &= \frac{{{{\rm{e}}^{ - 2\gamma t}}{J^2}\sin {{\left( {\chi t} \right)}^2}}}{{{\chi ^2}}},\\
{\rho _{11 }}\left( t \right) &={{\rm{e}}^{ - 2\gamma t}}{\left[ {\cos \left( {\chi t} \right) - \frac{{\gamma \sin \left( {\chi t} \right)}}{\chi }} \right]^2},
\end{aligned}
\end{equation}
where $\chi {\rm{ = }}\sqrt {{J^2} - {\gamma ^2}}$. The evolution of the post-selected occupation number ${{\rho _{11 }}}$ was fit to an exponentially decaying sine function to determine the coherence-decay rate and the Rabi oscillation frequency. These results were consistent with the direct theoretical approach for the evolution of the qubit wave function under $H_{eff}$.

\section{Derivation of evolution time} \label{evolution time}
The evolution operator of effective passive $\mathcal{PT}$-symmetric non-Hermitian Hamiltonian can be written as:
\begin{equation}
U = {e^{ - i{H_{eff}}t}}{\rm{ = }}\left( {\begin{array}{*{20}{c}}
{{{\rm{e}}^{ - t\gamma }}\left( {\cos \left[ {t\sqrt {{J^2} - {\gamma ^2}} } \right] + \frac{{\gamma \sin \left[ {t\sqrt {{J^2} - {\gamma ^2}} } \right]}}{{\sqrt {{J^2} - {\gamma ^2}} }}} \right)}&{ - \frac{{{\rm{i}}{{\rm{e}}^{ - t\gamma }}J\sin \left[ {t\sqrt {{J^2} - {\gamma ^2}} } \right]}}{{\sqrt { - {J^2} + {\gamma ^2}} }}}\\
{ - \frac{{{\rm{i}}{{\rm{e}}^{ - t\gamma }}J\sin \left[ {t\sqrt {{J^2} - {\gamma ^2}} } \right]}}{{\sqrt {{J^2} - {\gamma ^2}} }}}&{{{\rm{e}}^{ - t\gamma }}\left( {\cos \left[ {t\sqrt {{J^2} - {\gamma ^2}} } \right] - \frac{{\gamma \sin \left[ {t\sqrt {{J^2} - {\gamma ^2}} } \right]}}{{\sqrt {{J^2} - {\gamma ^2}} }}} \right)}
\end{array}} \right)
\end{equation}

To solve the passive PT-symmetric non-Hermitian Hamiltonian of the two-dimensional quantum
brachistochrone problem, one can choose the basis so that the initial and final states are given
by:$\left| {{\psi _i}} \right\rangle  = \left( {\begin{array}{*{20}{c}}0\\1\end{array}}
\right)$, $\left| {{\psi _f}} \right\rangle  = \left( {\begin{array}{*{20}{c}} a\\b\end{array}}
\right)$. Here, the system is initialized in the $|1\rangle$ state. The relation $\left| {{\psi _f}}
\right\rangle  = {e^{ - i{H_{eff}}t}}\left| {{\psi _i}} \right\rangle $ ($\hbar=1$) takes the form:
\begin{equation}
\left( {\begin{array}{*{20}{c}}
a\\b
\end{array}} \right) = \frac{{{{\rm{e}}^{ - t\gamma }}}}{{\cos \left[ {\arcsin \left[ {\frac{\gamma }{J}} \right]} \right]}}\left( {\begin{array}{*{20}{c}}
{ - {\rm{i}}\sin \left[ {t\sqrt {{J^2} - {\gamma ^2}} } \right]}\\
{\left( {\cos \left[ {t\sqrt {{J^2} - {\gamma ^2}}  + \arcsin \left[ {\frac{\gamma }{J}} \right]} \right]} \right)}
\end{array}} \right)
\label{ab}
\end{equation}
If $b=0$, the evolution time from $|1\rangle$ to $|0\rangle$ can be deduced:
\begin{equation}
t = \frac{{\pi  - 2\arcsin \left[ {\frac{\gamma }{J}} \right]}}{{2J\sqrt {1 - {{\left( {\frac{\gamma }{J}} \right)}^2}} }}
\end{equation}
Likewise, if the system is initialized in the $|0\rangle$, the evolution time from $|0\rangle$ to $|1\rangle$ becomes:
\begin{equation}
t = \frac{{\pi + 2\arcsin \left[ {\frac{\gamma }{J}} \right]}}{{2J\sqrt {1 - {{\left( {\frac{\gamma }{J}} \right)}^2}} }}
\label{S8}
\end{equation}

\section{The mapping of $\mathcal{PT}$-symmetric non-Hermitian Bloch sphere }\label{CPT123}

For a $\mathcal{PT}$-symmetric Hamiltonian which has specific phase $\phi_J$ in the off-diagonal terms:
\begin{equation}
H=\left(\begin{array}{cc}
	i \gamma & J e^{i\phi_J} \\
	J e^{-i\phi_J} & -i \gamma
\end{array}\right)
\label{H}
\end{equation}

By defining $\alpha\equiv\arcsin{\frac{\gamma}{J}}$, the eigenstates of Eq.(\ref{H}) become:
\begin{equation}
\left|\varepsilon_{+}\right\rangle=\sqrt{\frac{J}{\Delta E}}\left(\begin{array}{c}
e^{i\phi_J}e^{i\alpha/ 2} \\
e^{-i\alpha / 2}
\end{array}\right), \quad\left|\varepsilon_{-}\right\rangle=\sqrt{\frac{J}{\Delta E}}\left(\begin{array}{c}
i e^{i\phi_J}e^{-i\alpha/ 2} \\
-i e^{i\alpha/ 2}
\end{array}\right)
\label{eig}
\end{equation}
where $\Delta E\equiv2\sqrt{J^2-\gamma^2}$ is defined as the difference between two eigenvalues. State $|\psi\rangle=\cos \frac{\theta}{2}|0\rangle+e^{i \phi} \sin
\frac{\theta}{2}|1\rangle$ can be represented with the basis of $\left|\varepsilon_{+}\right\rangle$ and $\left|\varepsilon_{-}\right\rangle$:
\begin{equation}
|\psi\rangle=m_{1}(\theta, \phi ,\frac{\gamma}{J},\phi_J)\left|\varepsilon_{+}\right\rangle+m_{2}(\theta, \phi ,\frac{\gamma}{J},\phi_J)|\varepsilon_-\rangle
\end{equation}
where $m_{1}$ and $m_{2}$ are complex numbers
\begin{equation}
\begin{aligned}
m_{1}(\theta, \phi ,\frac{\gamma}{J},\phi_J)&=\sqrt{\frac{J}{\Delta E}}[e^{-\frac{i\alpha}{2}} e^{i \phi } \sin \left(\theta/2\right)+e^{\frac{i\alpha}{2}} e^{-i\phi_J}\cos \left(\theta/2\right)],\\
m_{2}(\theta, \phi ,\frac{\gamma}{J},\phi_J)&=\sqrt{\frac{J}{\Delta E}}[i e^{\frac{i\alpha}{2}} e^{i \phi } \sin \left(\theta/2\right)-i e^{-\frac{i\alpha}{2}}e^{-i\phi_J} \cos \left(\theta/2\right)].
\end{aligned}
\end{equation}
Thus, the state can be rewritten as:
\begin{equation}
|\psi\rangle=r_{1}e^{i\Phi_{1}}\left|\varepsilon_{+}\right\rangle+r_{2}e^{i\Phi_{2}}|\varepsilon_-\rangle
\label{phi}
\end{equation}
where
\begin{equation}
\begin{aligned}
r_1&=\sqrt{\frac{1+\sin (\theta ) \cos (\alpha-\phi -\phi_J )}{2 \cos (\alpha )}},\\
r_2&=\sqrt{\frac{1-\sin (\theta ) \cos (\alpha +\phi+\phi_J )}{2 \cos (\alpha )}}.
\end{aligned}
\end{equation}

As $|\psi\rangle$ in Eq.(\ref{phi}) is a four parameters vector, by extracting a common phase, $|\psi\rangle$ turns into:
\begin{equation}
|\psi\rangle^{\mathcal{CPT}}=R \cos \frac{\Theta}{2}\left|\varepsilon_{+}\right\rangle+R \sin \frac{\Theta}{2} e^{i \Phi}|\varepsilon_-\rangle
\label{Psi}
\end{equation}
where $R\equiv\sqrt{{r_1}^2+{r_2}^2}$, $\Phi\equiv\Phi_{2}-\Phi_{1}$ and
$\Theta\equiv2\arcsin(\sqrt{\frac{1-\sin (\theta ) \cos (\alpha +\phi+\phi_J )}{2+2 \sin (\alpha ) \sin
(\theta ) \sin (\phi+\phi_J )}})$. This helps us deriving the distribution of the state represented by Bloch sphere in
$\mathcal{CPT}$-conjugate inner product space. Moreover, the evolution operator $U_{H_{\mathcal{PT}}}=e^{-iH_{\mathcal{PT}}t/\hbar}$ become unitary by setting $R=\sqrt{\langle\psi_0|\psi_0\rangle^\mathcal{CPT}}$, where $|\psi_0\rangle$ is the initial state. In the main context, $R=\sqrt{\langle1|1\rangle^\mathcal{CPT}}$ because the system is initialized to $|1\rangle$. Consequently, a non-Hermitian Bloch sphere have been constructed.

\end{document}